# Electrostatic field acceleration of laser-driven ion bunch by using double layer thin foils

Xin Wang (王鑫)[1], Wei Yu (余玮)[2], and Edison Liang[1*]

1 Rice University, Houston, Texas 77005-1892, USA
2 Shanghai Institute of Optics and Fine Mechanics, Chinese Academy of Sciences, Shanghai 201800, China

Monoenergetic ion bunch generation and acceleration from double layer thin foil target irradiated by intense linearly polarized (LP) laser pulse is investigated using two-dimensional (2D) particle-in-cell (PIC) simulations. The low-Z ions in the front layer of the target are accelerated by the laser-driven hot electrons and penetrate through the high-Z ion layer to generate a quasi-monoenergetic ion bunch, and this bunch will continue to be accelerated by the quasi-stable electrostatic sheath field which is formed by the immobile high-Z ions and the hot electrons. This mechanism offers possibility to generate monoenergetic ion bunch without ultrahigh-contrast and ultrahigh gradient laser pulses in beam generation experiments, which is confirmed by our simulations.

Introduction

In recent years, the availability of ultra-short ultra-intense (USUI) laser pulse makes possible the development of compact laser-driven ion accelerators[1-3]. Short pulse laser plasma is becoming a very important component of high energy density physics. The resulting energetic ions have many important applications, ranging from medical proton therapy[4], diagnostics for laser-plasma interaction[5], and fast ignition in inertial-confinement fusion[6]. In most of these applications energetic ions of sufficiently high energy and brightness are required.

Several acceleration mechanisms have been proposed and demonstrated to generate high quality ion beams, for example, target normal sheath acceleration (TNSA)[7-10], radiation pressure acceleration (RPA)[11-15], Coulomb explosion[16], and the break-out afterburner (BOA)[17]. During the early years, TNSA is the main scheme for ion acceleration. When an USUI laser pulse irradiates a solid foil, the relativistic electrons generated in the laser-plasma interaction can easily penetrate though the foil and form an electron sheath on the rear side of the foil[18-21], the resulting space-charge field then pulls out plasma ions from the back surface. The energetic ions thus generated can be accelerated to several MeV[22-25]. With the rapid development of laser technology, such as plasma mirror technology[26], the RPA has drawn increased attention to generate relatively higher density and higher energy ion beams. But the instability suppression and beam confinement are still the intractable issues, especially for LP driven lasers.

In this paper, we introduce a novel mechanism to combine TNSA and radiation pressure (RP) to generate quasi-monoenergetic ion beams and high gradient electrostatic field, then the field gradually release its energy and accelerate the ion beams to higher energy without much expansion. We study the mechanism using two-dimensional particle-in-cell (PIC) simulations[27-28]. LP laser and double-foil target are used. LP laser is used for generating hotter electrons and double-layer target with low-Z (low density) plasma in front and high-Z (high density) plasma in the rear. This allows the pulse to penetrate the low-Z part but be blocked by the high-Z part. This kind of foil targets makes the laser energy absorption a little higher than laser pulse directly irradiating high-Z targets[29], and makes the conditions of laser beam increasing to peak intensity in several periods and circularly polarized laser not so crucial to generate high quality ion beams.

PIC simulation

We performed 2D3V (two dimensional in space and three dimensional in velocity) collisionless PIC simulations[27-28] on the interaction of LP Gaussian laser pulses with a double layer thin slab target. The target consists of two layers, low-Z plasmas on the front side and high-Z plasmas on the rear side. The Gaussian LP laser pulse is normally incident along the $z$ axis from the left vacuum region into the simulation box, and irradiates the target. Laser wavelength is $\lambda_0 = 1\mu m$.

Here we discuss the benchmark case where $a_L (= \frac{eE_L}{m\omega_0 c}) = 15$, and the spot size (diameter) is $d = 10\lambda_0$. The corresponding laser intensity is of order $3\times 10^{20} \times \sin^2(t\pi/2\tau_i)\times \exp(-4r^2/d^2) Wcm^{-2}$ (without plateau part). The laser temporal profile is a sine-squared distribution in time with total duration $\tau = 45T_0$ ($15T_0$ for ascend laser pulse, $15T_0$ for descend part, and $30T_0$ for plateau part in middle). Here $T_0$ is the laser period, which equals to $3.3 fs$. The laser parameters are close to the Texas Petwatt Laser parameters[30]. Thus, the electron motion is in the highly relativistic regime, and the laser ponderomotive force is dominant in the laser-plasma interaction. In our simulations, the low-Z plasma (hydrogen Z=1) density on the front size is $n_0 = 4n_c$ (alterable density in our scheme, but must be much lower than the backside plasma density), and we assume the high-Z plasmas (aluminum Z=13) has electron density $n_e = 650n_c$ and ion density of $n_i = 50n_c$. The two layers are both initially $0.1\lambda_0 (=0.1\mu m)$ in thickness, and located at $4.9 \leq z/\lambda_0 \leq 5.0$ and $5.0 < z/\lambda_0 \leq 5.1$, respectively, $n_c = 1.1\times 10^{21} cm^{-3}$ is the critical density.

In the simulation, the computation box is $30\lambda_0 \times 19.2\lambda_0$, the spatial mesh contains $3000\times 1920$ cells, and each cell contains 625 ions and 625 electrons. The electron-ion mass ratios are $1/1836$ for hydrogen plasmas and $1/49572$ for aluminum plasmas, respectively. The initial plasma electrons and ions are assumed to be fully ionized, with temperature $T_e = T_i = 1\ keV$. The time step of the simulation is $0.006T_0$. Absorbing boundary conditions are used for both the $y$ and $z$ simulation-box boundaries.

Simulation results and explanations

When the temperature increases in a plasma, the Coulomb cross section falls rapidly, and the collisions become rare. Here we quote the results of Wesson (2003)[31], for reference.

Electron collision time with plasma target:

$$\tau_e = \frac{3}{4\sqrt{2\pi}} \frac{m_e^{1/2}(k_B T_e)^{3/2}}{n_i Z_i^4 e^4 \ln \Lambda_e} = 1.09 \times 10^{-11} \frac{T_e^{3/2}}{n_i Z_i^2 \ln \Lambda_e} s,$$

Ion collision time with plasma target:

$$\tau_i = \frac{3}{4\sqrt{\pi}} \frac{m_i^{1/2}(k_B T_i)^{3/2}}{n_i Z_i^4 e^4 \ln \Lambda_i} = 6.60 \times 10^{-10} \frac{A^{1/2} T_i^{3/2}}{n_i Z_i^4 \ln \Lambda_i} s,$$

Here $T_e$, $T_i$, $n_i$, $Z_i$, and $A$ are the electron temperature in $keV$, ion temperature in $keV$, ion density, charge number and atomic number, respectively, and the $\ln \Lambda_e$, $\ln \Lambda_i$ relates to electron density and particle temperature, which are close to 10 in most of the conditions. We can get mean free path in unit of $\mu m$, $3270 \frac{T_e^{3/2}(1022 T_e + T_e^2)^{1/2}}{n_i Z_i^2 (511+T_e) \ln \Lambda_e} \mu m$ and

$1.98 \times 10^5 \frac{A^{1/2} T_i^{3/2}(1022 a T_i + T_i^2)^{1/2}}{n_i Z_i^4 (511a + T_e) \ln \Lambda_i} \mu m$, here $a = m_i / m_e = 1836$.

For simplicity, we assume the test particles (electron and ion) have the same temperature as the plasma target. We find that electrons with energy above 1.4 $keV$ can penetrate 0.1 micron aluminum plasma, and for hydrogen ion the energy is 35 $keV$. In our scheme, the particle will be heated quickly as explained below, so the collisionless PIC simulation is applicable.

As we expected, the double layer structure with different materials help us to generate quasi-monoenergetic hydrogen ions. Figure 1 shows the charge density distributions of two kinds of plasmas, solid blue line and dotted green line indicate the aluminum charge density and hydrogen charge density, respectively. Actually, the densities are indistinguishable. We just separate them for convenient observation. The charge densities are normalized by their own initial densities. As Figure 1 (a) shows, $1T_0$ later than laser pulse front reached the target, at $t = 6T_0$, the electrons of the hydrogen layer were pushed away from their initial position, some of them even past the aluminum layer. Due to the heavier mass, the protons were almost not affected by the lower laser intensity; and the electrons and ions of aluminum layer are also rarely affected, since the Al density was too high for the laser front. As time processes, the higher intensity part of the laser pulse reaches the target, and the two kinds of plasmas would be heated as Figure 1 (b) shows. At very close position of aluminum front surface, the most interesting phenomenon happened at $t = 15T_0$: the electrons from the hydrogen layer were heated quickly and expanded to large scale, but due to electrostatic effect, a part of the electrons were captured near the aluminum front surface (proton distribution as Fig 2 (b) shows); the proton bunch would be separated to two parts, and it formed a hole (low density well) that time; and the electrons and ions from aluminum were accumulated to higher density at that location.

For clarification, we diagnose the electric field distributions, as Figure 2 shows. We plot the results with the same times as Figure 1, at $t = 6T_0$ and $t = 15T_0$. Red solid line indicates the laser amplitude. The laser pulse is linearly polarized with $y$ axis direction. Here the laser amplitude is normalized by its initial peak amplitude; blue solid line indicates electric field along $z$ axis direction. And the black dashed line and the blue dashed line indicate the hydrogen layer initial front surface and the initial boundary of the two materials, respectively. And the proton and aluminum ions densities are represent by green and purple solid lines. In Figure 2 (a), it is easy to find that the laser field penetrated into the hydrogen layer because of the thickness is comparable to its skin depth, but the laser field was blocked by the aluminum layer due to the high density particles, at $t = 6T_0$, as we expected. The electron movement caused charge separation forming an accelerating field along $z$ axis near the two materials boundary. As time progressed, the heated electron pulled out the proton from the rear side of the hydrogen layer, similar to the TNSA process, and some parts of the protons penetrated into the aluminum layer. At the same time, the higher intensity laser irradiated the aluminum layer, and the RP process happened, and the aluminum electrons and ions accumulated to higher density. In Figure 2 (b), when at $t = 15T_0$, the higher intensity laser pulse reached the target, but the pulse still could not penetrate the aluminum layer, and was reflected. We find a sudden change of accelerating field (field along $z$ axis) near the aluminum front surface, which becomes positively changed and the protons separated as Figure 1 (b) and Figure 2 (b) showed. Then one part of the protons would penetrate the aluminum and then passed through aluminum layer, simultaneously, another part of the protons would be decelerated and then reaccelerated along $-z$ axis direction. After the protons were totally separated to two parts, a monokinetic proton bunch is generated.

In our scheme, we using quasi-stable electrostatic sheath field to continue accelerating the monokinetic proton bunch, instead of laser electric field, for instability suppression and maintaining high quality bunch for longer time. As Figure 3 (a) shows, the laser field was blocked and reflected by the aluminum layer. But during the process, the electrons from aluminum were heated and expanded out of the foil region. Due to the large ion electron mass ratio, the aluminum ions were not pulled out of the region. Hence the electrostatic field formed at the rear side of aluminum, as Figure 3 (b) shows. When the laser kept interacting with the aluminum foil, the laser energy continued changing to potential energy, then potential energy released as time progressed, and accelerated monokinetic proton bunches, which had passed the aluminum ion layer, to monoenergetic proton bunches, simultaneously. Although, the electrostatic field peak amplitude reaches just one fifth to one third of the laser field peak amplitude, it is easier to accelerate a collimated bunch with some initial velocity to higher velocity since the longer interaction time. And because the electrons expanded to large scale, the aluminum ions expanded very slowly, before the field came back to neutral, the protons were kept accelerating.

In Figure 4, we plot accelerating electrostatic field (along $z$ axis) and proton density together. Here the proton density is represented with arbitrary unit. The proton bunch was located in the gradient of accelerating field, which was formed by very slowly moving aluminum ions (solid black line indicates the aluminum ion density) and surrounding hot electrons. The Coulomb expansion process would take place during the proton bunch moving forwards, but the $z$ axis electrostatic field not only accelerated the bunch, but also suppressed the bunch expansion. We see that the forward moving proton bunch had asymmetric density distribution, left side is much higher than right side, (see right solid red lines in Figure 4). But for the backward moving proton bunch, due to the effect of coming laser pulse and the –z direction electrostatic field, the peak density is in the middle of the bunch (see left solid red line in Figure 4).

From Figure 4, we also find that there was almost no overlap between aluminum ions and proton bunches, at $t = 42T_0$. Actually, the proton bunch would totally pass the aluminum ions, and the gap between the two kinds of ions become larger and larger as time processes, due to Coulomb repulsion. But the most import thing is this scheme converts laser accelerating field to electrostatic accelerating field. Since the aluminum target prevented the laser to penetrate the target all along, the resulting electrostatic accelerating field is always pointing in the same direction with the bunch moving in the direction we want, and without any transverse components quiver field, much like the results which use circularly polarization laser to accelerate foil, suppressing the R-T like instability[32].

In figure 5, we show proton densities at different times. Figure 5 (a) and (b) are two dimensional proton density distributions at $t=42T_0$ and $t=78T_0$, respectively. Figure 5 (c) presents proton densities on the laser axis with interval of $12T_0$. The initial proton density is $n_p=4n_c$. It is easy to find that the average proton density is close to $0.05n_c$ at $t=90T_0$. The corresponding energy density is $5\times10^{14} J/m^3$. It is a very high energy density plasmas with many potential applications.

For more information of the high-quality proton bunch, we also diagnose the proton spectrum at different times. In the Figure 6 (a), the blue, green, red, cyan, purple, yellow, black and light green lines are indicating the proton spectrums at t=$6T_0$, $18T_0$, $30T_0$, $42T_0$, $54T_0$, $66T_0$, $78T_0$ and $90T_0$, respectively. We also calculate the average energy of the forward moving proton in the bunch. We find the field continued to accelerate the proton to average energy of 55 $MeV$ at $t=90T_0$. But the proton bunch had long low-energy tails as in Figure 5 (b). If we choose about 40% of the most forward protons, their average energy is about 65 $MeV$, at $t=90T_0$.

As we know, shape tailoring could help us to improve the bunch quality, but the easiest way is to cut off the tails of the bunch, as showed in Figure 5 (b), is to chop off the transverse diameter of the hydrogen shell. By selecting only protons within $1\mu m$ of the laser axis, the proton spectrum is improved to become more narrow-band as in Figure 7.

We mentioned above that energy density is an important indicator for high energy density physics. Here we also investigate how the target parameters affect the results. With the given laser condition, we find neither the aluminum layer nor the hydrogen layer could be too thick or too thin. If the aluminum is too thin, it allows the laser to penetrate the target and then cause R-T like instability. If the aluminum target is too thick, it blocks too many hot electrons, which causes fewer protons to be pulled over the aluminum. As for the hydrogen, if it is too thick hot electrons will pull more than one layer of protons over the aluminum layer. If it is too thin, on the other hand, there will be too few hot electrons to pull out enough protons.

In our scheme, the front layer density must be much lower than the rear one for high quality bunch generation. So we vary the density of the front layer to get more stable experimental conditions. We studied two cases, one with $0.05\lambda_0$ thick aluminum, $0.07\lambda_0$ thick hydrogen; another with $0.1\lambda_0$ thickness aluminum, and $0.1\lambda_0$ thickness hydrogen. We find that hydrogen densities from $4n_c$ to $10n_c$ give the optimal density for comparable stable high energy bunches, as Figure 8 shows. The first case gives asymptotic proton density about $0.028n_c$, average proton energy about $93 MeV$. The second case gives proton density about $0.046n_c$, average energy about $60 MeV$.

## Conclusion

In conclusion, we have investigated detailed high quality proton beam generation from a double-layer thin target irradiated by a high intensity laser pulse, by using 2D PIC simulations. The target is comprised of a low density hydrogen layer and a higher density aluminum layer. This special structure makes the driving laser not penetrate the Al target, and the generated protons in front penetrate the aluminum layer, then to be accelerated to quasi-monoenergetic proton bunch. The acceleration depends on converting laser field into electrostatic field and releasing electrostatic energy to accelerate protons. This suppresses the R-T like instability.

## Acknowledgment

This work was supported by DOE grant DE-SC-000-1481. Han Xu is strongly acknowledged for helpful discussions.

Fig 1. Net charge density distribution profiles along laser propagation axis ($y=0$) at (a) $t=6T_0$, (b) $t=15T_0$. Here $T_0$ is the laser period, and the densities are normalized by their initial densities respectively. The solid blue line and dotted green line indicate the charge densities for aluminum layer plasma and hydrogen layer plasma, respectively; the dashed red and dotted-dashed purple lines show the total charge and $E_z$, respectively. For clarity, they are moved away from the axis for +0.75 and -0.9, respectively.

Fig 2. $E_y$ (red line), $E_z$ (blue line), proton density (green line) and aluminum ion density (purple line) on the axis at (a) $t=6T_0$ (b) $t=15T_0$. Here the $E_y$ is normalized by initial laser amplitude. Black dashed line shows the front boundary of hydrogen layer, and the blue dashed line indicates the boundary of hydrogen layer and aluminum layer. And part of the zone in the middle is enlarged for clarity in the insets.

Fig 3. Contour plots of (a) Laser field $E_y$, at $t=42T_0$; (b) $E_z$, at $t=42T_0$.

Fig 4. Ez (blue solid line), aluminum ion (black solid line), and proton density profile at $y=0$ (red solid line), at $t=42T_0$.

Fig 5. Hydrogen ion density distribution at (a) $t=42T_0$ and (b) $t=78T_0$. (c) Hydrogen ion density profile along laser axis (y=0) at t=$6T_0$, $18T_0$, $30T_0$, $42T_0$, $54T_0$, $66T_0$, $78T_0$ and $90T_0$, respectively. The density is normalized by critical density $n_c$.

Fig 6. (a) Hydrogen ion energy spectrum at t=$6T_0$, $18T_0$, $30T_0$, $42T_0$, $54T_0$, $66T_0$, $78T_0$ and $90T_0$, respectively. (b) Average energy of forward moving proton (solid blue line) and average energy of the 40% protons which are moving faster.

Fig 7. Proton energy spectrum at t=$6T_0$, $18T_0$, $30T_0$, $42T_0$, $54T_0$, $66T_0$, $78T_0$ and $90T_0$, respectively.

Fig 8. Energy density at $t = 90T_0$ for two kinds of double layer target, blue line: aluminum thickness $0.05\lambda_0$, hydrogen thickness $0.07\lambda_0$; green line: aluminum thickness $0.1\lambda_0$, hydrogen thickness $0.1\lambda_0$.

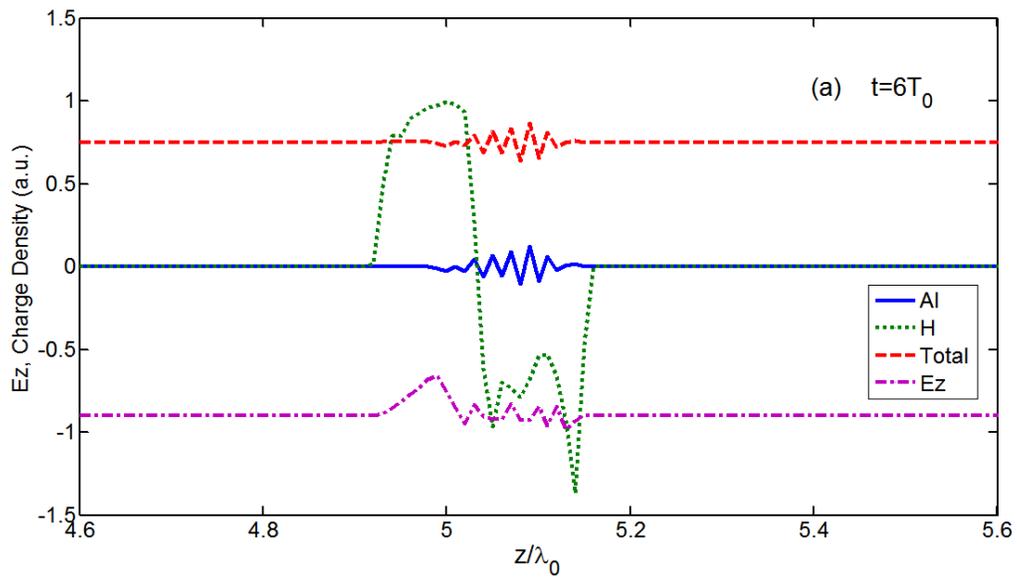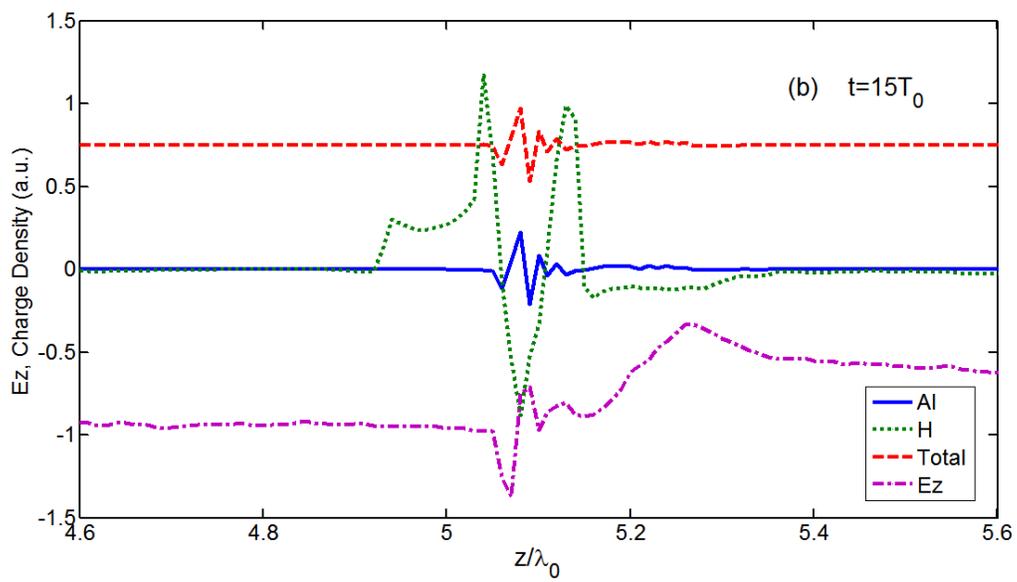

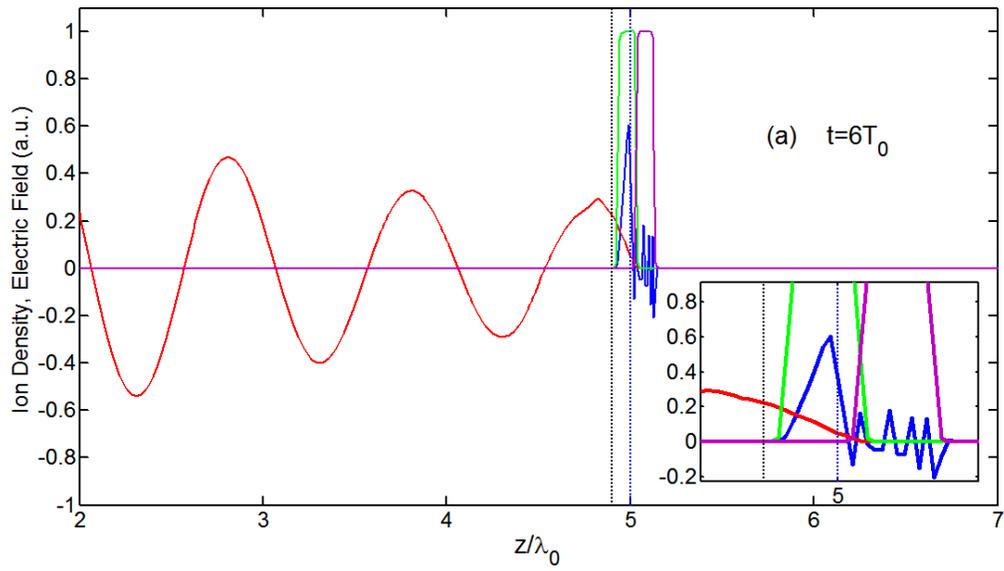
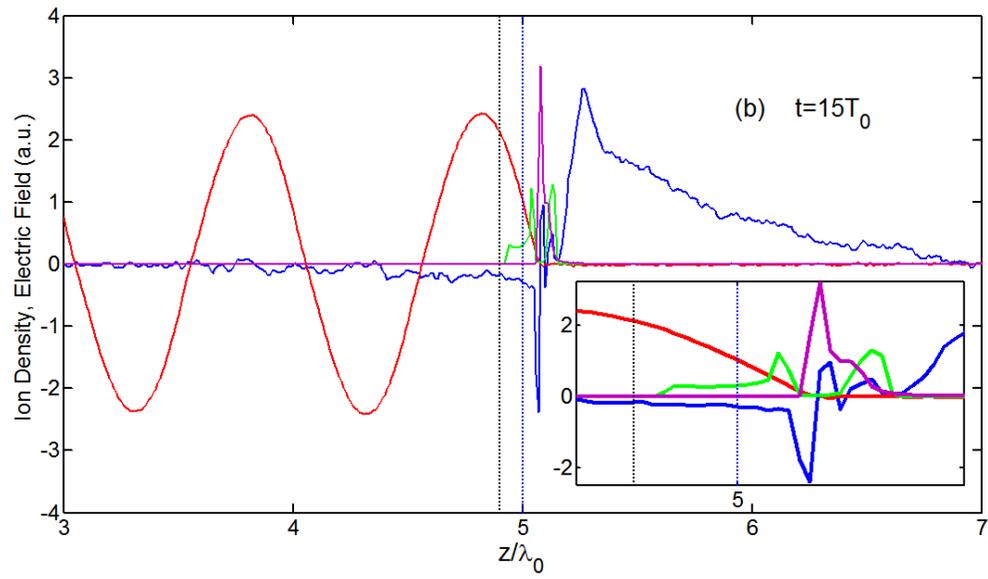

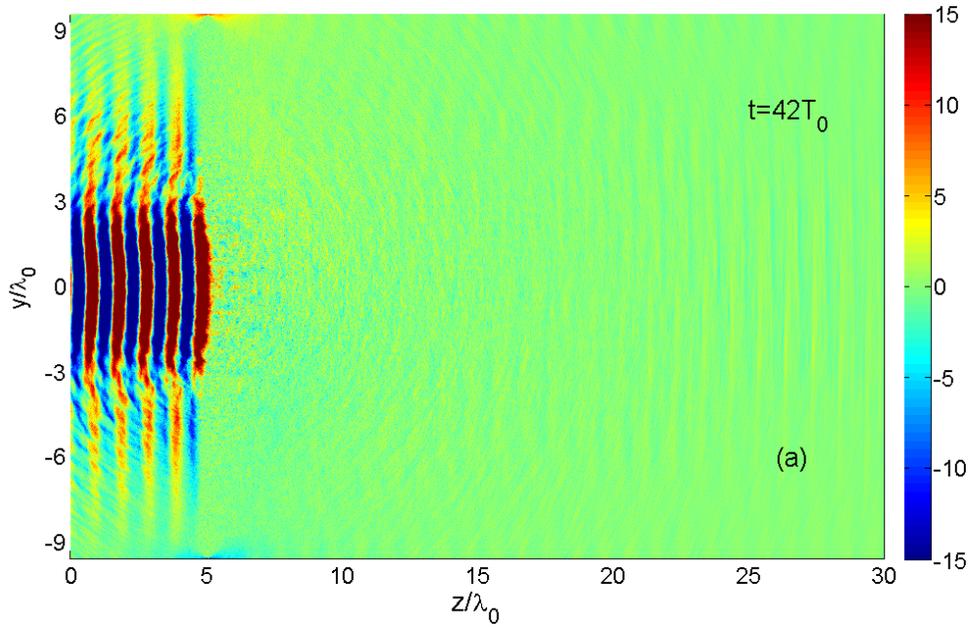

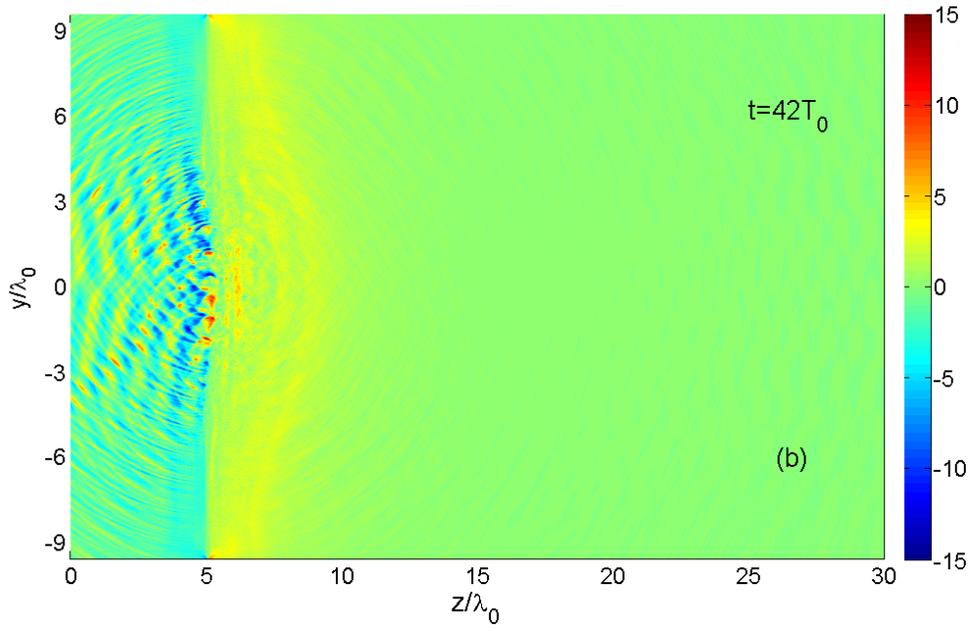

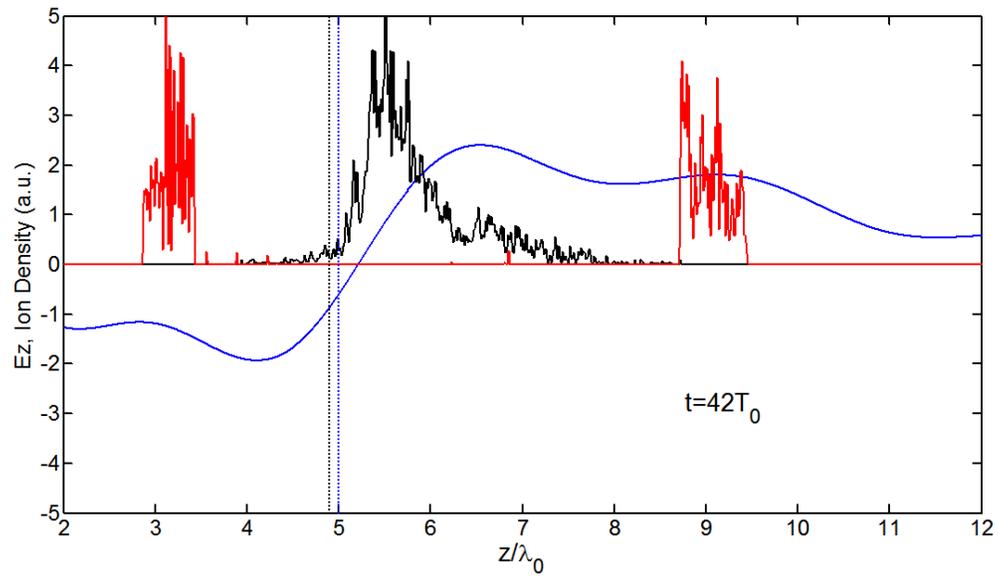

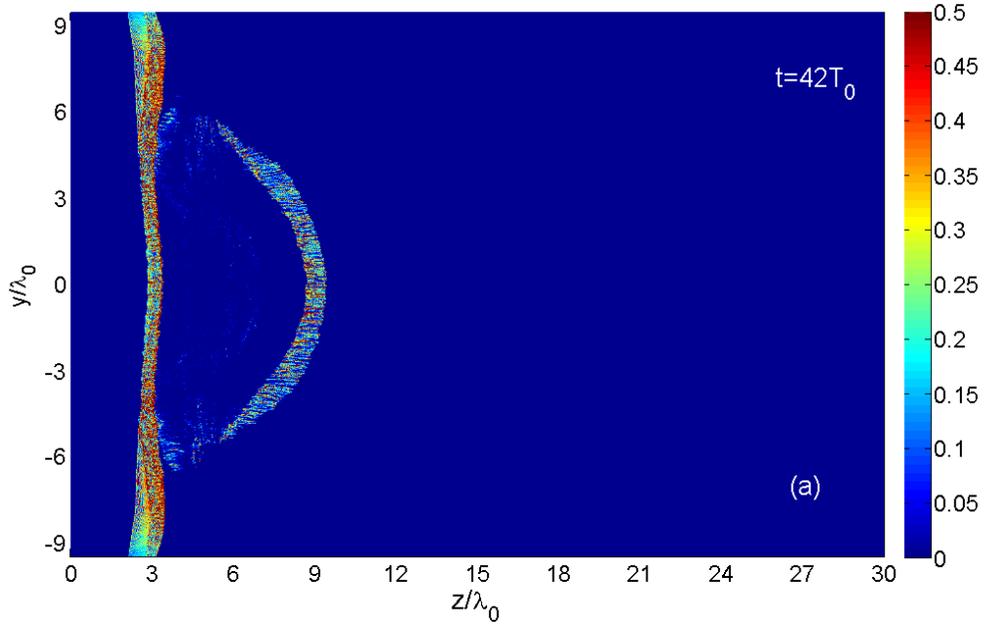
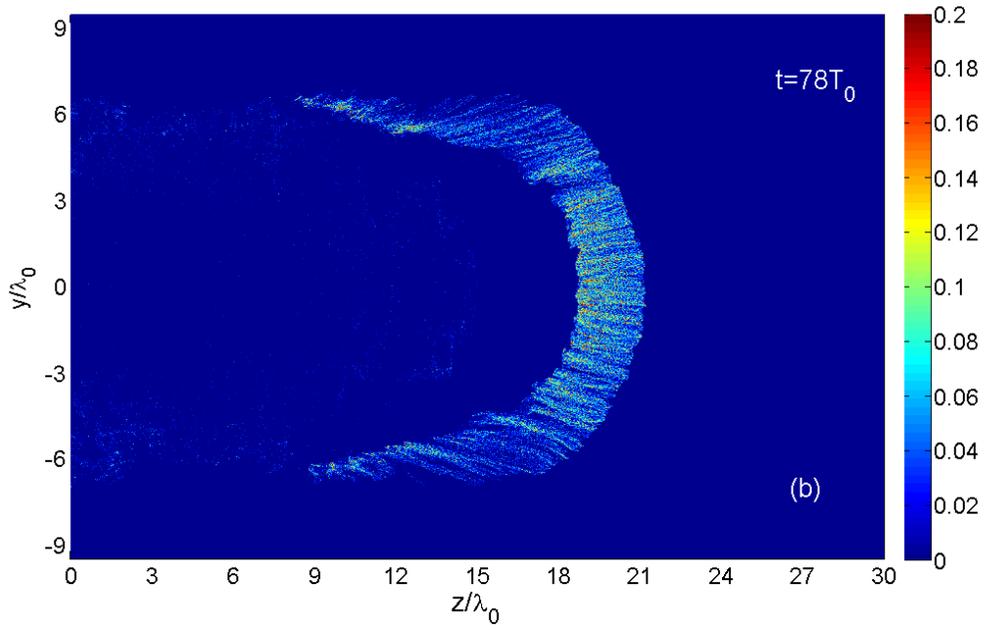
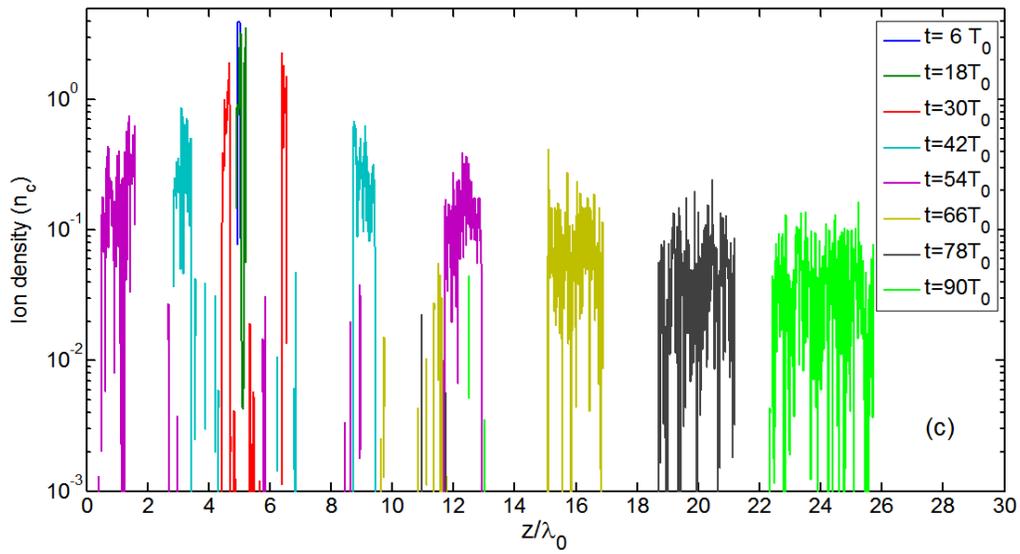

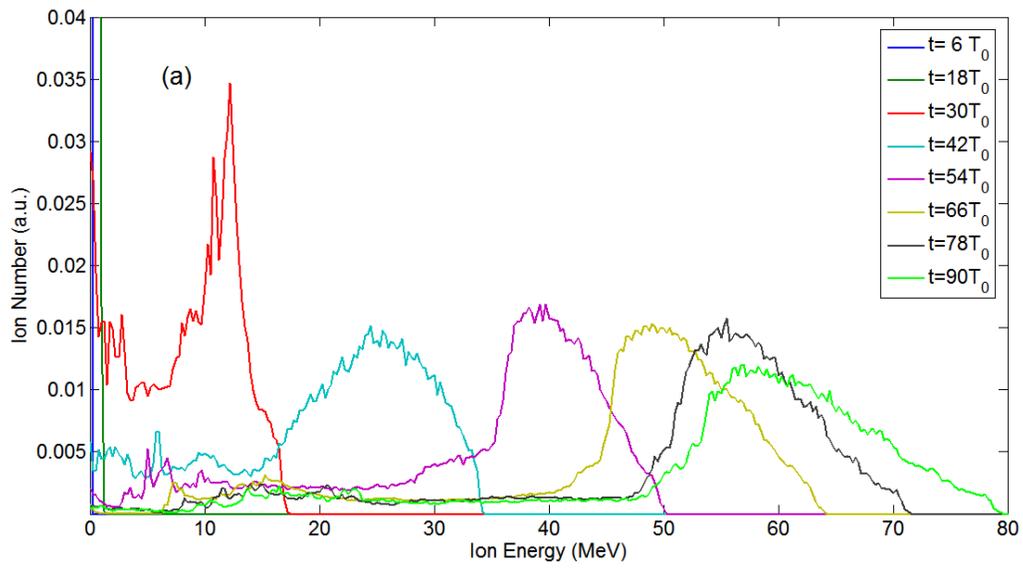
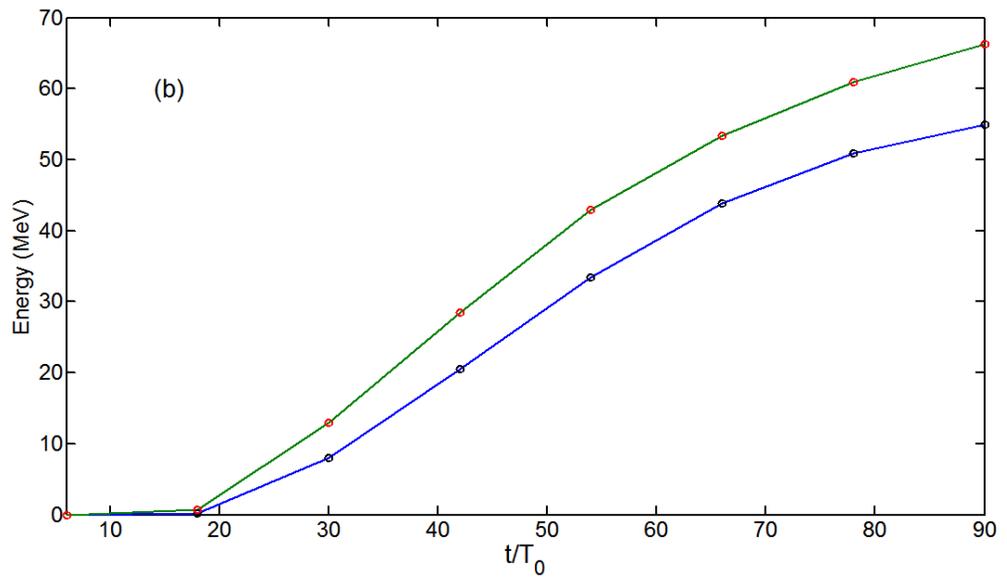

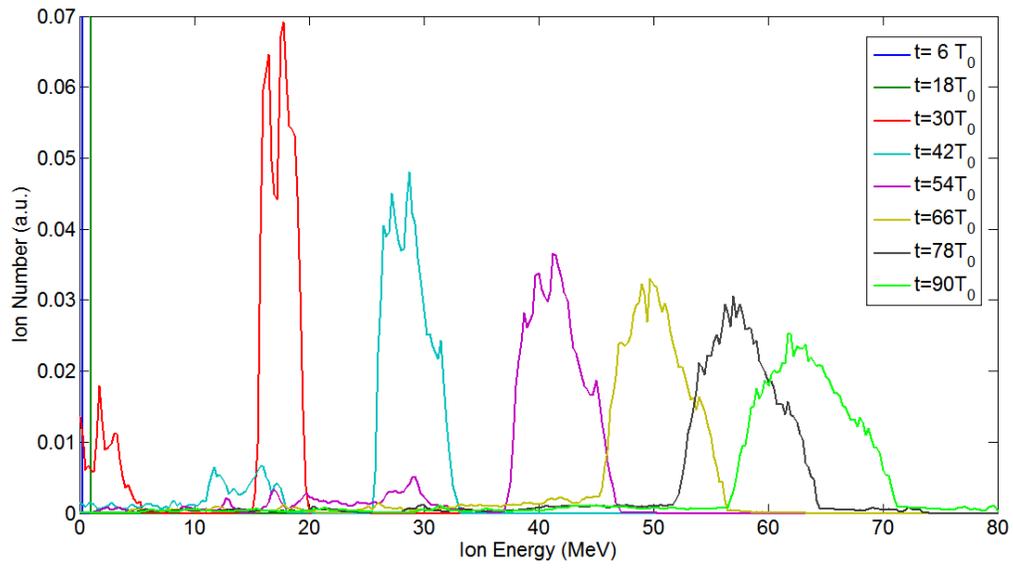

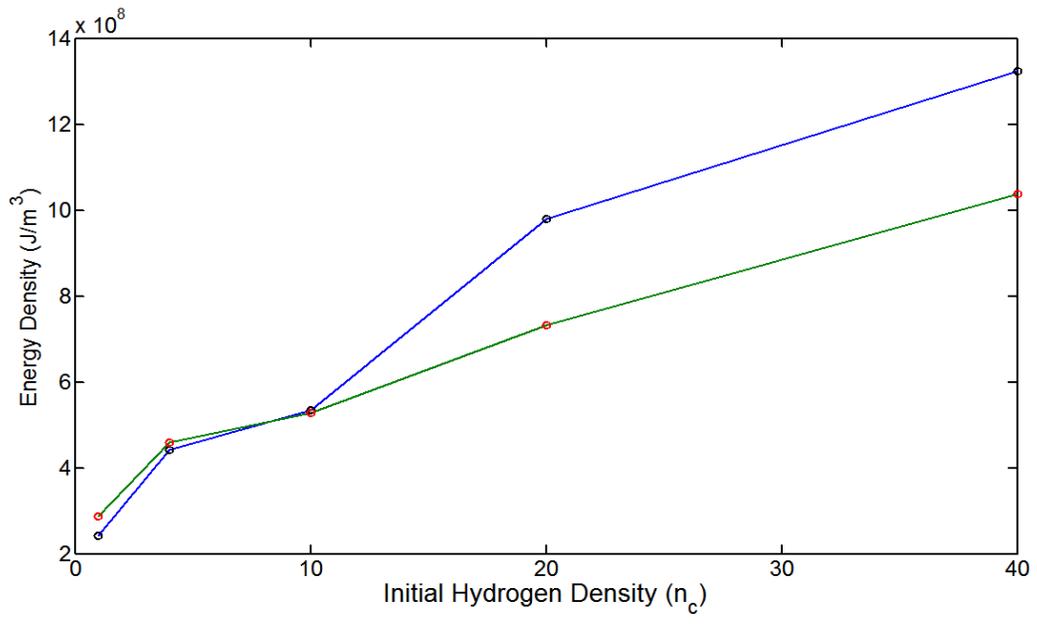